\documentclass[sigconf]{acmart}
\AtBeginDocument{%
  \providecommand\BibTeX{{%
    \normalfont B\kern-0.5em{\scshape i\kern-0.25em b}\kern-0.8em\TeX}}}

\setcopyright{acmcopyright}
\copyrightyear{2025}
\acmYear{2025}
\acmDOI{XXXXXXX.XXXXXXX}

\acmBooktitle{} 
\acmPrice{15.00}
\acmISBN{978-1-4503-XXXX-X/18/06}

\usepackage{graphicx}
\usepackage{float}
\usepackage{subcaption}
\usepackage{xcolor}

\newenvironment{quoteitalicized}
    {\begin{quote}}
    {\end{quote}}
\newcommand{\quotes}[2]{\begin{quoteitalicized}\textit{#1} {#2}\end{quoteitalicized}}

\begin{document}

\title{Embracing Transparency: A Study of Open Science Practices Among Early Career HCI Researchers}


\author{Tatiana Chakravorti}
\affiliation{%
  \institution{The Pennsylvania State University}
  \city{State College}
  \state{Pennsylvania}
  \country{USA}
  \postcode{16802}
}
\email{tfc5416@psu.edu}

\author{Sanjana Gautam}
\affiliation{%
  \institution{The Pennsylvania State University}
  \city{State College}
  \state{Pennsylvania}
  \country{USA}
  \postcode{16802}
}

\author{Sarah M. Rajtmajer}
\affiliation{%
  \institution{The Pennsylvania State University}
  \city{State College}
  \state{Pennsylvania}
  \country{USA}
  \postcode{16802}
}
\email{smr48@psu.edu}

\renewcommand{\shortauthors}{Trovato and Tobin, et al.}
\begin{abstract}
Many fields of science, including Human-Computer Interaction (HCI), have heightened introspection in the wake of concerns around reproducibility and replicability of published findings. Notably, in recent years the HCI community has worked to implement policy changes and mainstream open science practices. Our work investigates early-career HCI researchers' perceptions of open science and engagement with best practices through 18 semi-structured interviews. Our findings highlight key barriers to the widespread adoption of data and materials sharing, and preregistration, namely: lack of clear incentives; cultural resistance; limited training; time constraints; concerns about intellectual property; and data privacy issues. We observe that small changes at major conferences like CHI could meaningfully impact community norms. We offer recommendations to address these barriers and to promote transparency and openness in HCI. While these findings provide valuable and interesting insights about the open science practices by early career HCI researchers, their applicability is limited to the USA only. The interview study relies on self-reported data; therefore, it can be subject to biases like recall bias. Future studies will include the scope to expand HCI researchers from different levels of experience and different countries, allowing for more justifiable examples.
\end{abstract}





\maketitle

\section{Introduction}
In recent years, numerous scholars have raised concerns regarding a lack of transparency in Human-Computer Interaction (HCI) research, see e.g., \cite{wacharamanotham2020transparency, ballou2021you, aitamurto2019hci}. In particular, authors have highlighted lack of transparent reporting of both quantitative and qualitative methodologies \cite{aeschbach2021transparency, pater2021standardizing, talkad2020transparency, chakravorti2024reproducibility}, questionable statistical practices \cite{vornhagen2020statistical, cockburn2018hark}, and insufficient methodological rigor \cite{greenberg1992weak}. A recent survey study of CHI authors from 2018-2019 found data sharing to be very uncommon \cite{wacharamanotham2020transparency}. Another study revealed that while only 2\% of CHI papers shared their interview protocols in 2017, this statistic increased to 25\% by 2022 \cite{salehzadeh2023changes}. The same study reported raw data sharing rate of 7\% and processed data sharing rate of 17\% at CHI 2022 \cite{salehzadeh2023changes}. 

These are amongst the emphases of \emph{open science}, the broad set of practices and principles that strive for transparency, sharing, and inclusivity \cite{miguel2014promoting, bradley2020reducing, aguinis2019transparency, raff2023siren}. Beyond normative and philosophical motivations, e.g. \cite{merton1942note,anderson2010extending}, open science practices are closely tied to reproducibility, replicability--and ultimately--confidence in published findings \cite{miguel2014promoting, bradley2020reducing, aguinis2019transparency, raff2023siren}. The latter has particularly faltered since the last decade's string of disappointing large-scale replication projects across the social and behavioral sciences and beyond \cite{nosek349corresponding,camerer2016evaluating,camerer2018evaluating,errington2014open,willis2020trust,kapoor2023leakage}
In response to these revelations, open science communities have launched innovative solutions aimed at strengthening the entire research workflow, from conception and study design to data collection and analysis \cite{nosek2015promoting, nosek2016transparency, obels2020analysis}. These efforts have already had significant impacts on both individual and institutional levels, many of which are well-documented \cite{silverstein2024guide, mazarakis2020gamification}. For instance, the Special Interest Group on Computer-Human Interaction (SIGCHI) recommends providing supplementary materials for ACM publications to enhance replicability \cite{echtler2018open}. Within the ACM SIGCHI community, numerous research events \cite{bruckman2017cscw, fiesler2018research} have focused on discussions around research ethics. In the HCI field, the conversation on transparency has been highlighted through community-driven initiatives, such as RepliCHI \cite{wilson2014replichi, wilson2013replichi}, as well as various opinion pieces \cite{talkad2020transparency}. Additionally, the ACM  has introduced an artifact review and badging scheme to motivate these practices. However, the effectiveness of this scheme is debatable \cite{zong2023open}. 

In parallel, a separate thread of research has emerged developing methods to synthesize and evaluate existing literature to better determine which findings can be considered reliable and in which contexts \cite{yang2020estimating, pawel2020probabilistic, chakravorti2023prototype, wu2021predicting, chakravorti2023artificial}. However, HCI research is very diverse \cite{wobbrock2016research}.  In particular, not all research fits neatly into the hypothesis testing or experimental structure, for example, qualitative research. Research shows sharing artifacts such as interview protocols, processed data, and software would facilitate the evaluation and contextualization of those research works \cite{mcgrath2018data, waitzkin1990studying}. 
However, the extent to which early career HCI researchers are being trained and encouraged to adopt these practices, their perceptions of rewards, and challenges they face in doing so remain largely unexplored.  Our work aims to fill this gap. 
This study is scaffolded by the following research questions (RQs):

    \begin{itemize}
    \item\textbf{RQ1}: What are early-career HCI researchers' perceptions and experiences around open science practices? How do these experiences differ across qualitative vs. quantitative researchers?
    \item\textbf{RQ2}: What are the potential benefits and barriers to adopting open science practices according to HCI researchers?
    \item\textbf{RQ3}: What is the role of HCI conferences and institutions in improving open research practices? What are the training and incentives needed to adopt these practices? 
    \end{itemize}

We observe a diverse yet often incomplete understanding of open science practices among early career HCI practitioners. The majority do not have an accurate and complete understanding of open science. Some researchers perceive open science as being primarily relevant to quantitative research, with less applicability to qualitative research methods. A percentage of the practitioner population conflates it with open publishing practices solely. A significant number of participants report limited knowledge and training in open science practices, particularly in how to effectively share data, protocols, and other research materials. Cultural resistance and lack of clear incentives emerge as major barriers for the advancement of open science practices. Our findings bring to light the critical role of conferences in driving these practices. Participants indicated that if major HCI conferences strongly encouraged open science practices, this would set a standard for the field and likely increase participation. Finally, we provide recommendations for the stakeholders across the scientific landscape in HCI to integrate these practices into the community.

\section{Related Work}
Our work builds upon and contributes to the literature on research ethics and open science practices, as well as transparency in HCI research. For all of these areas, we have focused on the literature of the HCI community and beyond. 

\subsection{HCI Community Engagement with Research Ethics}

For a community as rich and diverse as HCI, it poses a significant challenge to come to a shared understanding of what defines the research ethics concerning the field. For these reasons, ethics in HCI has attracted increasing attention in recent years as evidenced by a series of town halls \cite{bjorn2018research, frauenberger2017research, munteanu2019sigchi}, workshops \cite{davis2015ethical, waycott2015ethical, waycott2016ethical, fiesler2018research, densmore2020research, fleischmann2019good}, conference papers \cite{munteanu2015situational, fiesler2016exploring, wisniewski2017whose}, tutorials \cite{alzughbi2024epic}, panels \cite{talkad2020transparency}, special issues \cite{nathan2016disruptions}, journal articles \cite{benford2015ethical} and book chapters \cite{bruckman2014research}. SIGCHI community discussions in the past have raised questions surrounding the : (i) best practices to handle human-generated data, (ii) ethics of representation (especially within the cases of minority communities), and (iii) balancing of privacy and reproducibility among others \cite{fiesler2022research}. To further add to the tensions, the method of close association with participants adds an emotional perspective of ethics. There are studies that highlights the importance of emotion in socially-oriented HCI research and suggests the need for more reflexivity in the ethical review process \cite{hodge2020relational}. When we bring in the perspectives from HCI adjacent community efforts, for example, the Menlo Report \cite{dittrich2012menlo} identifies factors such as the rapid pace and decentralized nature of ICT, which create distance between researchers and research subjects, and increase the potential for harm due to the ease of data engagement. Another community that has voiced their ethical concerns, the Association of Internet Researchers \cite{franzke2022association}, focused on issues like consent, risks to researchers, and power imbalances, particularly in academic and industry partnerships. The discussions both within and in adjacent communities have led to some remarkable progress when it comes to defining research ethics in HCI. In the following sections, we discuss what these practices look like and further explore their adoption in the research world. 

\subsection{Open Science Practices in HCI}

Open science is an inclusive approach that promotes making scientific knowledge accessible and reusable for all, enhancing collaboration, and involving societal actors across all disciplines and practices \cite{das2021unesco}. To discuss the practices of open science, it might be important to outline the different points of the research life cycle and how open science practice applies at each phase. At the very inception of a research proposal, we have preregistration. Preregistering your research involves outlining your research plan before conducting your study and submitting it to a registry. The primary goal is to clearly distinguish between what you intended to do (confirmation) and what was discovered during the process (exploration)\footnote{https://www.cos.io/blog/preregistration-plan-not-prison}. Both aspects are crucial to scientific research, but mixing them up can lead to misunderstandings and misinterpretations of the findings. Preregistration helps maintain the integrity of your research by preventing self-deception and ensuring that the conclusions drawn are meaningful and transparent. Now to turn the attention to the next part of ethical participant recruitment and onboarding. A central argument made by \cite{dym2020ethical, klassen2022black, klassen2022isn, payne2023ethically} in studies on the ethical use of online trace data from marginalized groups is that “marginalization” should not be approached with a one-size-fits-all methodology. Instead, this research emphasizes the importance of adapting research practices to account for the unique context, history, and needs of each marginalized group \cite{payne2023ethically}. When talking about participation, it is critical to discuss consent from the participants. There are different aspects of the consent process that remain to be explored, for example, sharing data with industry collaborators \cite{cummings2021need}, online research participation \cite{zong2022bartleby}, and ongoing affirmative consent within sexual studies \cite{strengers2021can} among others. Next, open data, protocols, materials, and code foster transparency, reproducibility, and collaboration by making research content freely accessible, allowing others to review, replicate, and expand upon the research. Past work \cite{salehzadeh2023changes} has found that only 57\% of papers adhered to consent form practices, and rates for transparency, such as sharing interview protocols and justifying sample sizes, were even lower. With the review process, the goals of transparent peer review and open evaluation are to boost engagement and accountability in the peer review process and throughout the research lifecycle, encourage a more open and collaborative research culture, and strengthen trust in the scientific process. Finally, the growing evidence \cite{mckiernan2016open} highlights that openly sharing articles, code, and data benefits researchers by enhancing visibility, citations, and career opportunities, supported by increasing funder policies, academic recognition, and user-friendly tools.

\subsection{Transparency in HCI}
Numerous guidelines for reporting research methods are evidence of the importance of research method transparency \cite{nosek2015promoting}. To further analyze transparency related practices, we discuss qualitative and quantitative methods separately as they often require different treatment. Transparent research practices, crucial for reproducibility \cite{patil2016statistical} and replicability \cite{patil2016statistical}, involve detailed disclosure of methods, data, and other research artifacts. In HCI, replication studies are rare, leading to concerns about the field's research culture. While transparency in quantitative research focuses on method and data \cite{wacharamanotham2020transparency}, in qualitative research \cite{o2014standards}, transparency emphasizes method over data, with additional complexity around the concept of "translucency." Past works \cite{cairns2007hci, vornhagen2020statistical} have highlighted the importance of the continued need for improved transparency and accuracy in both quantitative and qualitative research reporting within HCI. However, previous studies on research ethics, openness, and transparency in HCI have not provided a comprehensive overview of these practices. 

\section{Methods}
We conducted 18 semi-structured interviews \cite{varanasi2022feeling, thakkar2022machine} with early career HCI researchers to understand their perceptions, experiences adopting open science in their research, and challenges of these practices. 

\subsection{Participant Recruitment}
For our study, we utilized direct emails, and social media platforms such as LinkedIn and Twitter to share our recruitment email and reach a diverse population of respondents. The target population was early-career HCI researchers such as PhD students, Post-doctorate, Assistant professors, and industry researchers. We recruited 18 participants (P1-P18) including 11 females and 7 males; 1 postdoc, 1 assistant professor, 2 industry researchers, and 14 PhD students. The age range of the participants was 25-35 years. The recruitment email and the advertisement contained information on the study, expected interview length, compensation, inclusion criteria, and a link to the screener survey (Google form). This survey helped us to select a diverse research group; from different professional backgrounds, different universities, different HCI research backgrounds, and genders. Three different time slots were provided by the participants for the semi-structured interview from those we selected one slot for each of them.

\subsection{Interview Protocol}
Semi-structured interviews \cite{sadek2023trends} were conducted virtually via Zoom video conferencing during July 2024. Interviews lasted between 40 minutes - 1 hour, depending on the length of participant responses to interview questions. All interviews were recorded and transcripts for analysis were generated through transcription capabilities native to Zoom. We asked participants to share their primary research method, their understanding of open science/open research practices, the conferences and journals in which they submit their research, their engagement in practices associated with open research, their familiarity with pre-registration, the importance of open science, ethical considerations to adopt these practices, support, and incentive needed to foster the open research culture. 

\subsection{Data Analysis}
We analyzed the interview transcripts using thematic analysis as outlined by Blandford \cite{blandford2016qualitative}. This qualitative data analysis method involves thoroughly reading the transcripts to identify patterns across the dataset and derive themes related to the research questions. We employed a collaborative and iterative coding process \cite{ding2022uploaders, huang2020you}. Initially, the first author read the interview transcripts multiple times to become familiar with the data. Open coding was then conducted to identify initial codes by the first author. These codes were subsequently organized into themes relevant to the primary research questions. Throughout the process, the first two authors met periodically to discuss the meanings, similarities, and differences of the identified themes and their relevance to the research questions. The author team made final decisions regarding the retention, removal, or reorganization of these themes collectively during weekly discussions. We came out of five themes, they are "familiarity with open science practices", "experiences with open science practices", "potential benefits from open science", "barriers adopting open science practices", and "Incentives and Recognition Required to Motivate". 

\subsection{Ethical Approval}
IRB approval for human subjects research was obtained before participant recruitment. Participants were fully informed about the nature of the study, potential risks, and their right to withdraw at any time without penalty before the study began. Consent was obtained without coercion. Data was stored securely and only used for agreed-upon purposes. Our work directly addresses research integrity and transparency. The main aim was to seed more inclusive conversations around research integrity and highlight the perspectives of the HCI community.

\begin{table*}[h]
\caption{Gender, rank, and affiliation of interview participants.}
\label{tab:subject1}
\begin{tabular}{|l|l|l|l|l|}
\hline
\multicolumn{1}{|c|}{\textbf{ID}} & \multicolumn{1}{|c|}{\textbf{Gender}} & \multicolumn{1}{|c|}{\textbf{University}} & \multicolumn{1}{|c|}{\textbf{Profession}} & \multicolumn{1}{|c|}{\textbf{Research field}} \\ \hline
P1  & Female & Academia     & Doctorate Candidate & Qualitative Method\\
P2  & Female   & Academia    & Doctorate Candidate  & Qualitative Method \\
P3  & Male & Industry     & UX Researcher & Mixed Method \\
P4  & Female &  Academia    & Doctorate Candidate & Qualitative Method\\
P5  & Male   &  Academia    & Doctorate Candidate & Mixed Method \\
P6  & Male & Academia   & Doctorate Candidate   & Mixed Method  \\
P7  & Female &  Academia  & Doctorate Candidate & Mixed Method  \\
P8  & Female & Academia  & Postdoc Candidate  & Mixed Method  \\
P9  & Male   & Academia  & Doctorate Candidate  & Mixed Method  \\
P10 & Female &  Academia    & Doctorate Candidate  & Qualitative Method \\
P11 & Female &  Academia   & Doctorate Candidate  & Mixed Method   \\
P12 & Male   &  Academia    & Doctorate Candidate  & Mixed Method  \\
P13 & Female &  Academia  & Doctorate Candidate  & Mixed Method  \\
P14 & Male   &  Academia   & Doctorate Candidate   & Qualitative Method        \\
P15 & Female & Industry    & UX Researcher & Qualitative Method \\
P16 & Female   & Academia     & Assistant Professor & Mixed Method  \\
P17 & Female   & Academia     & Doctorate Candidate  & Qualitative Method  \\
P18 & Male   & Academia   & Doctorate Candidate  & Mixed Method \\
\hline
\end{tabular}
\end{table*}

\section{Findings}

\subsection{Familiarity with Open Science Practices}
Here, we discuss participants' background knowledge and attitudes toward open science practices. This helps us contextualize their perspectives on adopting open science practices. 

\subsubsection{Perception about Open Science}
It was interesting to observe that while some participants were well-informed about open science, others were only partially familiar with it. Notably, only participant P17 mentioned she was not familiar with open science. Although many researchers had some level of familiarity with open science, their understanding of these practices varied. For example, researcher P4 described open science primarily as an open-access archive, where papers are freely accessible.

\quotes{I guess, according to what I know, open research practice is basically making research available to the community without paywalls, or membership. So I would say, the open-access archive is probably the biggest open research thing that I know.}{-P4}

Other researchers, such as P2, P13, and P14, also viewed open science primarily as open access, similar to P4. P14 specifically highlighted the cultural differences in open science practices across different countries. He also emphasized the importance of equity that open access can provide to developing countries.

\quotes{My understanding, is that open science is a worldwide movement in order to push the boundaries of knowledge to the people of 3rd World countries, with very limited access. With this, they will be able to read through the research articles that everyone does. Basically bringing equity.}{-P14}

Interestingly, we found that participant P10 (qualitative researcher) believed that open science practices are relevant only to quantitative researchers—a view also shared by other researchers, such as P1, P15, P4, and P13. According to P10, data sharing applies exclusively to quantitative work. 

\quotes{I am not so much familiar. But I have heard and read of things, but because I've not really done that much quantitative work I'm not familiar with the open research practices, I am on the qualitative side of things.}{-P10}

Despite the advancement of the open science movements and promoting best research practices, we found that researchers have very diverse perceptions of open science. For example, participant P11, a current PhD student, has a different understanding of open science practices compared to others. According to her, open science involves pre-registering and testing your hypothesis before publication, but she does not have proper clarity.   

\quotes{I'm currently doing a survey and before the data collection we need to do the preregistration. We need to test the hypothesis that I have heard from my advisor. I have not done it before.}{-P11}

Familiarity with open science or open research practices often depends on researchers' backgrounds, lab cultures, and collaborators. For example, P12 mentioned that he was introduced to these practices during his master's studies while working at the university library. P2, P6, and P7 mentioned their advisors motivate them to adopt some of these practices in their research. P15, an early career UX researcher from the industry, noted that most funding bodies are funded by taxpayer money, and therefore the public should have access to the findings and research materials. However, she also believes that open science does not apply to qualitative research and mentioned that her company does not allow the sharing of such materials.

\quotes{I think open science makes your research available to others without being behind a paywall, allowing not only scientists but anyone to access and learn about the findings. Since funding often comes from the public, why shouldn't the public, who have contributed to those funds, be able to see and use the research?}{-P15}

However, we observed P3, P6, P7, P8, P9, P16, and P18 are having a much better understanding of open research practices compared to the other participants. 

\quotes{I know, this concept, and I know it is related to making your research publicly available. And so it can be more reproducible and transparent. Such as preregistering, like research plan, and publishing your data set and code.}{-P18}


\subsubsection{Familiarity with Pre-registration}
We asked all participants if they were familiar with pre-registration and invited them to share their experiences if any. We observed that except for P8 and P16, all the other participants had never done pre-registration. However, the majority of the participants have heard about it except, for P3, P4, P13, and P17. Except for these participants the remaining researchers provided their understanding. For example, Participant P1, a 2nd year PhD student was not very sure about it but provided us with her basic knowledge. 

\quotes{It sounds familiar. But I might be wrong. It is that when you give your method first, and then they have to approve it, and after that only you can continue your data collection.}{-P1}

According to P18, researchers need to register for all the research plans which will be publicly accessible, and after preregistration, you can not alter your research method. 
We observed that according to the researchers, pre-registration is not required for qualitative research, it is for quantitative research, as mentioned by most of them. For qualitative research which is totally exploratory and subjective there, pre-registration doesn't make a lot of sense. Also, the usefulness is not clear enough for qualitative work according to the participants.

\quotes{I have heard about this somewhere like people preregister who are into quantitative research. I've heard them reporting their methods and all the kinds of analysis and tests even before collecting the data. But I am not sure how that is useful for qualitative studies.}{-P6}

Participants P5, P7, P8, P9, P10, P11, and P12 specifically mentioned that pre-registration is only required when a research study involves a hypothesis. P7 added that pre-registration is not a common practice in the HCI community. P9 further noted that, while there are benefits to having a timestamped version of the research plans before starting a study, even without hypotheses, but this alone is not a sufficient incentive for conducting pre-registration.

\quotes{I have heard about preregistration in one of my courses. I haven't seen that happen usually in HCI. So didn't have any experience with the preregistration process. Mainly it is more common for the projects that have hypothesis.}{-P7}

Another significant observation is that most HCI researchers submit papers to well-known top HCI conferences like CHI, and CSCW. Conference scale publication doesn't require a pre-registration for the study. It completely depends on the researcher if he feels the requirement or not. For example, we can clearly see it from P14.

\quotes{I have heard about it in one of my courses. I have never done this because our major outlets are Main CHI, CHI Play, and CSCW. I don't think that they require greater disruption.}{-P14}

P16 mentioned she did pre-registration because it was required for the journal submission. Participant P8 reported that she was asked by collaborators to pre-register her study as it was a collaboration with psychologists and psychology highly values open science. She also added that this was not for a conference submission but for a journal.  

\subsection{Experiences with Open Science Practices}
In this section, we asked researchers about their experiences with open science practices both as authors and reviewers for HCI conferences and journals. We observed that the majority had submitted to and reviewed for CHI, the leading HCI conference globally, with other conferences such as CSCW and CHI Play also being mentioned. Only a few researchers mentioned submitting to academic journals.

\subsubsection{Experience as Author}
This section represents the open science practices of these researchers as authors during the submission of their papers. All of them who have been involved in interview studies mentioned that they have never shared interview transcripts, as doing so could reveal personal details. The transcripts contain a significant amount of sensitive information, and it is not a common practice to share. If they need to share those lots of time and effort are needed to anonymize those transcripts which does not carry any incentives according to P1. Additionally, the top conferences do not have any mandatory regulations to submit these documents, and reviewers do not typically ask for them mentioned by Participant P1. For the follow-up questions, she mentioned that she did not provide the codebook generated from the transcript and the interview protocols as well during submission.

\quotes{For the qualitative paper, I submitted to CHI as a 1st author for that one I didn't provide any transcripts of my interviews. That's the only data that I collected. And I guess recordings. But that's not something that we would share because it would reveal the person's identity. It's not the common practice to share it. So it wasn't something that was requested, reviewers didn't even ask to see the transcripts.}{-P1}

Most HCI researchers do not consider the codebook as data that can be shared. However, we found that some researchers, like P2, P6, and P18, shared their interview protocols during conference paper submissions. P2 never shared her codebook. She also mentioned that she hasn't seen others sharing it either. P2 further noted that when she first saw that CHI submissions required making the paper open-access, she thought it was a very positive step. She believes this is beneficial for researchers who cannot access papers due to paywalls. Although she supports these practices, she is also willing to share the final outcomes of her project as open-source software. 

\quotes{For the CHI submission, they let you ask, like some additional documentation as supplemental. I've submitted my interview questions for that. But I haven't submitted the codebook. I wasn't sure but I don't think I've seen other people do it before. However, I'm working on a project that will result in something like a system. And I intend to share it as open source through Github.}{-P2}

Not only P1 and P2, but all other researchers (Except P6 and P18) mentioned that they do not share their codebook for the qualitative research because they do not see how it would be useful to others. They also noted that if the top conferences do not ask to share the codebook, which means it is generally not considered important to do so. However, qualitative researchers like P4 expressed concerns about sharing her work before the acceptance of the paper. 

\quotes{I feel like I wouldn't be open to sharing protocols and my data until the papers have been accepted. Because I don't want to make my process available to other people as it is my work.}{-P4}

Only, P6 and P18 mentioned that they have shared the codebook with the submission to provide more transparency and validity. They also believe that this can improve their acceptance of the paper. P5, a mixed method researcher mentioned they always try to share their code and data.  

\quotes{Yes, we have shared our code. We made the prototype public and we shared it publicly. The URL was public. We made it like a web-based prototype. So anyone can go to the URL and use that prototype to program.}{-P5}

Whereas, another mixed method researcher P8, mentioned she did not share the code because it was a prototype and it is not a common practice or asked to share these things. But for some journal submissions when it is mandatory to share the code and data, people share more mentioned by P16 and P8 because that is required to get the publication done. 

\quotes{The study during my PhD, we didn't share because we developed a prototype. We didn't make that code available and then the data for the interviews, I don't believe we made that available either.}{-P8}

Participants P1, P2, P4, P10, and P17 also mentioned that for the conference submission, they were not asked by the reviewers to share the code, data, or protocols. It is more about how you have described the method section clearly for the readers. The industry UX researcher P15 mentioned not sharing the data because the data is for the company and the company does not agree to share openly. However when we had a discussion with P3, another industry researcher mentioned data sharing, code sharing, and positive support for open research practices. 

\quotes{The product design that I did last year for CHI submission. We interviewed a bunch of product designers. I didn't share that, because that was like kind of internal data for the Company.}{-P15}

\subsubsection{Experience as Reviewer}
During the interview study, we also asked participants to share their experiences as a reviewer. The goal was to understand do they look for the supplemental materials submitted by the authors for the paper, and the importance of open sharing for peer review. The majority of the participants except P4, P6, P7, P9, and P12 have reviewed HCI papers. The majority of the participants have reviewed for CHI, CSCW, and some other NLP conferences. It was observed that the authors weren't asked by the participants to share code or data as it is not mandatory by the conferences. The participants try to evaluate the method section carefully to understand the clarity if that is not clear then they ask clarification questions as mentioned by P1.

\quotes{I have reviewed CSCW and CHI, but I never asked the authors to submit the data, codebook, or interview protocols. I feel like if the conference is not asking then why should I ask to share? I focus on the methods section if that has been explained clearly or not. If the method section is not clear then I ask my clarification questions.}{-P1}

The same responses were observed from the majority of the participants(P2, P3, P8, P10, P11, P17, P18). If it is not mandatory to share by the conferences then they usually don't ask to share during the review. Exceptionally, participant P5 mentioned that he has reviewed one such paper where they did not mention the parameters clearly which they have asked during the interviews. Therefore he asked to share the interview questions to have more clarity. 

\quotes{I also reviewed one such paper, where the authors have used different parameters during the interview but they haven't mentioned that clearly in the method section. Therefore I asked them to share the interview questions. If that's not there, it's very difficult to understand.}{-P5}

Participant P16, assistant professor of HCI, mentioned that is very important to share the protocols without which you cannot have the whole understanding of the method. She also mentioned that some of the papers share very unique methods but they don't share the protocols therefore other researchers can not implement them correctly. A very unique fact she shared about her experience as a reviewer for the CHI. Many authors started their experimental protocols during the review process for the CHI conference. But they drop the supplemental during the camera-ready version. 

\quotes{I have experience as a reviewer during the process of CHI, and lots of authors have started submitting the protocols. But during the camera-ready version, they just drop the supplemental.}{-P16}

Participant P7 mentioned clearly that in her lab the other researchers are from NLP, AI, and machine learning. Many of those AI and ML conferences have mandatory rules for data. code and materials sharing. Therefore researchers share more to get their paper published. From the observations, it is very clear that participants are really not very comfortable asking the authors of the papers during the peer review to share any data, code, or other materials as it is not mandatory by the conference itself. Especially the early career researchers when they have very little experience and they don't know how to ask these questions as they are also not involved that much in open sharing mentioned by P10. 

\subsection{Potential Benefits from Open Science}
The majority of the participants think open science is helpful in many ways. We observed there are few researchers who think there is no proper benefit of open research practices in HCI. For example, P4, a PhD candidate, mentioned during the interview that she doesn't know how open science can be beneficial for qualitative researchers like her. She also mentioned that it is more important for quantitative researchers. However, some of them have benefited from open science. For example, mentioned by Participant P1, as she is a qualitative researcher, does not have experience in survey design. For one of her studies, she had to do a survey she looked for open survey data sets and how the other researchers conducted the survey design. Many of the researchers have put their whole survey in the appendix with other materials. This helps her to design her study. 

\quotes{I was building a survey for my study. I did look for published papers that had the data set or like, had those appendices on, how they conducted the survey, and their survey design. They usually put their whole survey in the Appendix. That helped me build my own survey because I had never done it before.}{-P1}

According to P6, a PhD candidate, open science practices can increase the citation count of the papers and also increase the chances of acceptance as the reviewers will feel the work is validated and rigorous. Therefore he is very positive about open science practices. 

\quotes{I'm all for sharing materials. Because number one, it increases the probability of acceptance. Because of the rigor of the paper right where a reviewer is reviewing it. They will feel more confident about the paper as it's transparent. Also in the future, it will help us to get more citation counts as well.}{-P6}

We found open source data is another thing that is helping many researchers. For example, participant P13 mentioned how open-source data sharing is helping her during the PhD journey. 

\quotes{For my research, I am using some clinical data that data was shared by others. That is some kind of open-source data which I am using for my research.}{-P13}

On the other hand, P11, a mixed method researcher, mentioned how she is using open-source code for her research which saves a lot more time than doing the same thing that already exists. Rather that time can be used to build something different than the existing work she mentioned. 

\quotes{I am using many open-source codes provided by the other researchers. This makes my research faster and I don't have to waste my time on what already exists in literature. Rather I can invest in something new.}{-P11}

The main benefits of open science mentioned by the participants are transparency and validity of research practices. It is ethical to ask to have transparent research mentioned by P17. These are the basic practices of good research and these increase the reproducibility and replication mentioned by P18. However, he mentioned that reproducibility is not for qualitative researchers rather this is for quantitative research. Due to the increased use of technology and large language models, like ChatGPT, it is very easy to fake data and make fake research mentioned by P3, P5, P6, and P7. Open research practices can reduce these unethical practices and improve the research quality.

\quotes{I feel people can also fake data. And it's very easy nowadays, right? Because of chatGPT and other tools like that. You can always create a normal-looking research but which is fake.}{-P3}

A very important point has been mentioned by P14, an international PhD student, the importance of open sharing of research papers in the global south mainly for those universities that can not pay for access to publications like IEEE or ACM. Open science has opened the doors for them at least to access a lot of research papers that were completely impossible or they had to use pirated versions in the past. 

\quotes{The university where I got my bachelor's and master's degree back in my home country, did not have any subscription services to access research papers. We could not afford to buy a single piece of paper for $10$ dollar. I know it's not a good way to describe it. But we unfortunately had to pirate the papers, because there was essentially no other way for us.}{-P14}

\subsection{Barriers to Adopt Open Science Practices}
\subsubsection{Lack of Incentive}
Academic reward systems in the HCI community often prioritize publications and citations over open practices, leading researchers to focus on these metrics rather than sharing data, code, methods, or any other materials mentioned by all the participants. Highlighted by all the participants that currently there is no incentive to do open research practices which needs more effort and time. Also, the industry participants feel the same. Some researchers are highly motivated by themselves for example P6 but that number is very rare. Currently, it completely depends on personal ethics and self-motivation as mentioned by P8.

\quotes{Currently there is no incentive for open science practices, it is all about self-motivation, ethics, and discipline.}{-P8}

\subsubsection{Cultural Resistance}
Many scientific communities are motivated by traditional practices of closed research, where sharing data, code, protocols, and methods is not the norm. The majority of these participants mentioned that in HCI open sharing and open research practices are not common. P1 and P17 are concerned that openly sharing data or methodologies will lead to more criticism or negative scrutiny during the peer review process. 

\quotes{I am scared to share the interview protocols because I feel this can create more criticism by the reviewers. Open research practices are not a common practice in the HCI community.}{-P17}

The primary focus of all participants is clear: they aim to get their papers published. Their priority is to publish in top HCI conferences, such as CHI. For these conferences, it is not required to submit data and code, which diminishes the emphasis on open research practices. When is it not required that implies it is not that important to submit according to the majority of the participants except P2. We observed these practices are very culturally diverse depending on the lab culture or the research community. For example, P17 added that last year 2023 their lab group started talking about all these practices. 

\quotes{From last year I remember our lab members and my advisors started to talk about data sharing in the CHI and CSCW communities. I heard that researchers started to emphasize the importance of sharing their interview protocols and other materials in the supplemental.}{-P17}

On the other hand side, P7 mentioned that her co-authors don't support open sharing which highlights the traditional lab culture. 

\quotes{My co-authors mentioned to me that the norm in the field of HCI is not to share qualitative data. If we ask participants that we are going to share all of their data, like everything that they say in the interview, in conferences or journals. They might not be very transparent with us,}{-P7}

As we observed in the HCI community, open science practices are not yet widely adopted. As a result, researchers are skeptical about the usefulness of sharing their materials. They are uncertain whether, after investing significant time and effort, others will actually notice or utilize their shared resources. 

\subsubsection{Limited Knowledge and training}
We observed from this interview study that researchers lack the necessary training and support to implement open science practices effectively, from data management, data sharing, and protocol sharing to navigating open-access publication. For example, Participant P10 has mentioned clearly that she doesn't know how to share the codebook. The codebook has different levels of coding and which one should be shared is not clear to her. Also, P2 mentioned that most of the time codebooks are not very well structured which can be shared. The semi-structured interviews, do not always have the exact questions in the protocols, they always go with the flow and explore more insights. P1 and P10 mentioned that they don't know in which structure they should share these protocols as they are semi-structured. Also, participants are not trained to anonymize the transcripts.

\quotes{We did rounds of open coding and deliberation. Then we sort of like focus those together. So we did some rounds of focus coding. And finally, we divided all of these relevant codes into the 3 themes, that sort of answer our research questions. So I'm just trying to wonder, at what stage would the code book require to be shared? I'm not sure if I know of the methodology, for at least semi-structured interviews to share transcripts and the protocols.}{-P10}

It was very interesting to observe that Participant P4, a first-year PhD student in HCI, doesn't know where to share these materials. After more in-depth conversations she mentioned that she never thought the supplemental is for data sharing or protocol sharing. 

\quotes{At least for the papers that I have submitted to the CHI conference, there isn't really like on Pcs. I haven't really seen anything to make these kinds of submissions for open research practices.}{-P4}

A significant number of researchers report having no formal training or attending workshops related to open science practices. The lack of structured training is a major barrier to adoption. Some participants like P1, P5, P7, P11, and P12 mentioned that even when training was available, it was very basic and lacked depth, particularly in areas like reproducibility, and transparency. There is a clear need for more formalized, detailed training programs on open science practices mentioned by the participants. This training should be integrated into existing curricula and made available to researchers at all career stages, especially early-career researchers. Participants suggest that such training would be particularly useful if it provides practical guidance on how to implement open science practices in their specific research areas.

\subsubsection{Time and Effort}
Preparing data and methods for open sharing requires a significant amount of time and effort which itself is very difficult for early career researchers. The majority of these participants like P1, P5, P8, and P11 mentioned that they do not have that much time to invest in open research practices. For example, P11 mentioned that she does not have time to manually go through all transcripts and anonymize them. And, if she is going to invest that much time and effort then she would expect some incentive for that which is not currently afforded in the academic system. 

\quotes{I don't have time to manually go over every transcript and make them anonymized. It will take lots of effort without any incentives. It is not required in this community.}{-P11}

For qualitative research, sharing transcripts is a major challenge. It involves considerable personal data and, at the same time, it is not well structured as it is automatically coded. In addition, researchers do not always code the entire transcript as it would take considerable time.

\quotes{If I have to share the transcripts then I have to fix the whole transcript so it's very time-consuming. I only fix the relevant parts that are needed. Also if no one is doing that in your community then why should I invest that much time which is not cost-effective.}{-P1}

\subsubsection{Intellectual Property}
Concerns about losing control over data and intellectual property can make researchers reluctant to share their work openly which was observed from the conversations with P4, P8, and P14. These three participants are very scared that their ideas and data can be scooped or stolen by others. According to the P4, she is not comfortable with the sharing because it is her own research and she wants to continue doing based on her findings in the future. P8, the postdoctoral researcher also thinks in the same way and mentions her concern about the protection of intellectual property.  

\quotes{I think the only reason to withhold sharing is the risk for sharing, and I think that my biggest concern with open science practices is, where are the protections for intellectual property? I think that there's like a novelty aspect to HCI research where the most novel ideas gain the most attraction. And when you share your research, and then someone else goes and uses what you've built, then it reduces the novelty of your subsequent work.}{-P8}

\subsection{Incentives and Recognition Required to Motivate}
One of the significant barriers to open science practices is that there is no incentive for researchers. Researchers need a clear understanding of the benefits associated with adopting open science practices as mentioned by P1. She added a need for cost-benefit analysis which justifies the additional time and effort required to make data, code, and other materials openly available. Many researchers like P4, and P8 currently see no incentive or direct benefit from sharing their data, which discourages them from engaging in open science. Conferences and academic institutions could incentivize open science by offering awards for papers that share supplementary materials or demonstrate reproducibility mentioned by P3. 

\quotes{I think conferences can provide different awards or recognition for papers that share more supplementary materials, and which are in general very open and reproducible, that will be a good way to encourage and recognize their efforts.}{-P3}

This recognition could encourage more researchers to engage in open science practices. The adoption of open science practices would be more widespread if major conferences and journals required or strongly encouraged some of these practices. For instance, if leading conferences in the HCI field, such as CHI, mandated open science practices, it could set a standard that others would follow, as mentioned by P7. 

\quotes{I think that it may be beneficial if the major conferences like CHI require you to practice open science, then most HCI researchers will be inclined to do that, a lot of the conferences in machine learning started doing that when a lot of papers came out which mentioned that they are not being able to replicate.}{-P7}


\section{Discussion}
The findings from our study highlight several key factors surrounding the adoption of open science practices within the HCI research community. We observed sharing is not a very common practice in HCI \cite{wacharamanotham2020transparency}. As we all know, there are important ethical reasons to make our research as open as possible while ensuring it remains appropriately restricted when necessary. One significant reason is that approximately 80\% of the University’s funding for research and teaching comes from public sources. Therefore, it is only fair to make as much of our work accessible to the public who ultimately supports it. Additionally, we can never predict how these intermediate research outputs might be utilized. In this discussion, we explore the implications of our findings and propose recommendations aimed at addressing the challenges associated with the adoption of open science practices within the HCI community. These recommendations are designed to mitigate existing barriers and facilitate a smoother integration of open science principles into the community's research workflows.

\subsection{Limited Understanding of Open Science}
Our interviews reveal a wide range of understandings and perceptions of open science practices among early career HCI researchers. While some participants are cultivated in the principles of transparency, some only think about reproducibility and others have a more limited or even narrow knowledge of what open science leads to, often equating it solely with open-access publications. Some participants expressed concerns about privacy and security in data sharing, even though UNESCO's recommendation on open science emphasizes the importance of sharing while safeguarding privacy and security \cite{unesco2021unesco}. Researchers don't have proper knowledge about preregistration and its usefulness. This suggests a need for clearer communication and education around the full spectrum of open science practices, including data sharing, code sharing, pre-registration, and the open sharing of research protocols \cite{schmidt2016stepping, beaudry2024incoming, button2020grassroots}. Interestingly, some researchers perceive open science as being primarily relevant to quantitative research, with less applicability to qualitative research methods. There is always the debate that qualitative research is subjective and therefore not reproducible \cite{aguinis2019transparency, l2019preregistering}. For quantitative research, it has been encouraged to do open science more often but for qualitative research, researchers have mixed reactions\cite{chauvette2019open}. HCI researchers mentioned that open science is not a common practice in their community. However, the demand to share qualitative data and interview protocols is becoming more prevalent due to the extensive use of large language models. Open science is not just for reproducibility or replicability rather, it is much bigger than that, it is for transparency and equity as well \cite{seo2017equality, fleming2021open}. This perception reflects a significant challenge in promoting open science across all areas of HCI.

\subsection{Potential Benefits and Challenges}
Given that 98.4 \% of works at CHI involve human participants, open science practices emerge as furthermore, important \cite{salehzadeh2023changes}. The potential benefits of open science are recognized by many participants, particularly in terms of increasing the validity, transparency, and reproducibility of research \cite{corti2016editorial, powers2019open}. Open science practices also have the potential to democratize access to research, which is especially important for researchers in under-resourced regions like Global South who may not have access to subscription-based journals. However, in the last few years, research transparency has been improved in the HCI community. For example, research transparency in CHI has been improved from 2017 to 2022, especially where both qualitative data (by 13\%) and quantitative data (by 8\%) sharing improved \cite{salehzadeh2023changes}. Many of these participants mentioned the small change in the CHI conference for example free availability of their papers made them think about open science. Some of the participants have benefited from open data and code sharing which eventually saves their time for repetitive and established work. Rather they can invest that time to invent something new. Participants have mentioned that it would be helpful if authors could share their interview protocols or research protocols. While open science is increasingly recognized as important there are significant barriers and varied perceptions that hold its widespread adoption \cite{komninos2022value, ballou2021you, colusso2019translational}.  

The specific character of best practices, of course, varies across disciplines \cite{talkad2020transparency}. Many fields are still working to establish their own norms inspired by open science ideals \cite{errington2021challenges, delios2022examining}. Qualitative research often involves sensitive, context-dependent data, raising legitimate concerns about participant confidentiality and the appropriateness of sharing data like interview transcripts. Researchers have identified several methodological, legal, and ethical implications related to open data in qualitative research\cite{childs2014opening}. Through the interviews, participants voiced valid concerns regarding the challenges of sharing transcripts, particularly due to the need for thorough anonymization, which demands considerable time, effort, and specialized training. However, there can be situations when de-identified data may be asked to be shared by the funding agencies to get the funding. For example, Rebecca \cite{campbell2023open} and her group provides 32 qualitative interviews with sexual assault survivors' case examples to demonstrate how blurring and redaction techniques can be used to protect sensitive information. Cultural resistance also plays a significant role in hindering the adoption of open science in HCI. We observed that during the peer review authors don't share materials as it is not mandatory by the conferences which implies not important to share. The majority of the authors don't get comments about not sharing any materials from the reviewers, therefore they follow the same. Many researchers are concerned about potential criticism and scrutiny if they share their data and methods openly. This is further complicated by the nature of the research conducted within the community, where the emphasis on transparency and reproducibility may not always align with the challenges or priorities faced in practice. Overcoming these cultural barriers requires a shift in norms, where open science practices are not only accepted but expected as part of rigorous research. 

\subsection{Limited Knowledge and Training}
A significant number of participants reported limited knowledge and training in open science practices, particularly in how to effectively share data, protocols, and other research materials\cite{paret2022survey, maumet2022open, burleyson2021ten}. This lack of training is a critical barrier, as it leaves researchers unsure of how to implement these practices or navigate the associated challenges, such as anonymizing qualitative data or structuring codebooks for sharing. Participants mentioned not having any proper courses related to open science practices. Some of them added that the information they have from the institutions is little, not sufficient enough. Most of them have not even come across workshops or seminars on open science, but they often get other workshops like grant writing. 

\subsection{Limited Incentives and Recognition}
The lack of clear incentives is a major barrier to the adoption of open science practices in the HCI community \cite{manco2022landscape, chakravorti2024reproducibility}. Additionally, there is a lack of recognition for the significant additional effort and time investment required to undertake such work. Researchers expressed skepticism about the utility and recognition of sharing materials, particularly given the additional time and effort required. Without tangible benefits, such as enhanced career advancement opportunities or institutional recognition, there is little motivation for researchers to engage in these practices. Moreover, the academic reward system continues to prioritize traditional metrics like publications and citations over contributions to open science, further discouraging researchers. Early career researchers, particularly PhD students and Postdocs, are often pressed for time and if there is no proper incentive then it is very difficult to motivate them for these practices. Simple rewards like badges have some measurable impact on engagement with open science practices \cite{rowhani2020did, rowhani2018badges}. However, it has been highlighted that providing badges is not enough as a proper incentive for the participants. 

\begin{table*}[h]
\caption{Our Findings and Corresponding Recommendations}
\label{tab:finvsrec}
\begin{tabular}{|l|l|l|}
\hline
  & \multicolumn{1}{c|}{Main Findings}                                                                                                                                                                                                                    & \multicolumn{1}{c|}{Recommendations}                                                                                                                                                                                                        \\ \hline
1 & \begin{tabular}[c]{@{}l@{}}Lack of clear incentives is a major \\ barrier to adopting open science practices\end{tabular}                                                                                                                        & \begin{tabular}[c]{@{}l@{}}Institutions should recognize and reward \\ efforts in data sharing, reproducibility, \\ and open contributions to motivate researchers \\ and set quality benchmarks.\\Integrating open science contributions into the criteria \\ for promotions, selection, and tenure\end{tabular}                              \\ \hline
2 & \begin{tabular}[c]{@{}l@{}}A significant number of participants reported \\ limited knowledge and training in open science \\ practices, particularly in how to effectively \\ share data, protocols, and other research materials.\end{tabular} & \begin{tabular}[c]{@{}l@{}}Formalized training programs integrated into the \\ HCI curriculum, covering open science practices \\ like data management, ethics, and anonymization,\\ all components of open science. \\ especially for early-career researchers.\end{tabular} \\ \hline
3 & \begin{tabular}[c]{@{}l@{}}Barriers to participation in open science practices \\ due to the nature of data collection and methods.\end{tabular}                                                                                                 & \begin{tabular}[c]{@{}l@{}}A shift towards more open practices would benefit \\ the community and needs to come from top down.\end{tabular}                                                                                                 \\ \hline
4 & \begin{tabular}[c]{@{}l@{}}Lack of recognition mechanisms \\ for effective adoption of open science practices.\end{tabular}                                                                                                           & \begin{tabular}[c]{@{}l@{}}Encouraging OS practices at major HCI conferences \\ (like CHI and CSCW) could set a standard and increase \\ participation in the field. \\ Awards should be integrated by conferences.\\ \end{tabular}                                                           \\ \hline
\end{tabular}
\end{table*}

\subsection{Recommendation}
\subsubsection{Encourage a Cultural Shift Towards Openness}
Cultural resistance to open science practices is deeply rooted in the HCI community, with many researchers adhering to traditional norms of closed research\cite{azevedo2022towards}. To foster a cultural shift towards openness, senior researchers and leaders within the HCI community should advocate for open science practices, setting an example for others to follow. By publicly endorsing and practicing open science, these leaders can influence the broader community and help normalize these practices. Organize discussions, panels, and forums within the HCI community to engage researchers in conversations about the benefits and challenges of open science. These events should be inclusive, encouraging participation from researchers at all career stages and across different subfields of HCI. Establish peer support networks where researchers can share their experiences, challenges, and successes with open science practices. These networks can provide a platform for exchanging ideas and strategies, helping to build a community of practice around open science. The main obstacles to change are not technical or financial but social. Although scientists tend to maintain the status quo, they are also the ones who can drive change.

\subsubsection{Enhance Training and Education}
There is a clear need for more formalized training programs that are integrated into the educational curriculum for HCI researchers. These programs should be comprehensive, covering the practical aspects of open science, including data management, ethical considerations, and the use of open-access platforms\cite{Drach2022OPENSI, parsons2022community}. This should cover how the researchers can share protocols and anonymize qualitative data. Such training would be especially beneficial for early-career researchers, who are still forming their research practices and could be more easily guided toward a culture of openness. It is very important to know about the licenses that open science provides. For example, researchers can use open licenses, such as Creative Commons (CC) licenses, to control how their work is used. By strategically using open licenses, patents, defensive publications, and collaborative models, researchers and institutions can protect their intellectual property while contributing to the collective advancement of science.

Academic institutions and professional organizations should regularly offer workshops and seminars focused on the practical aspects and training of implementing open science practices for undergraduate and graduate students\cite{ignat2021built}. Many academic institutions have research policies or codes of practice that clarify the principles, ethical foundations, and expectations for researchers' behavior within the institution. However, open science policies, such as those related to open access and open data, are seldom included in these research policies\cite{lyon2016transparency}. Some institutions have begun developing, adopting, and implementing open science policies, primarily over the past few years \cite{kretser2019scientific}. Furthermore, the majority of conferences offer "Course" sessions as part of their registration, aimed at helping early-career researchers learn about emerging topics in the field. These sessions could also serve as an effective platform for disseminating knowledge and fostering discussions on open science practices. 

\subsubsection{The Role of Conferences and Journals}
The findings also suggest that the role of conferences and journals is pivotal in driving the adoption of open science practices. Participants indicated that if major HCI conferences like CHI mandated or strongly encouraged open science practices, this would set a standard for the field and likely increase participation. We understand there can be potential barriers to mandating these practices in the conferences. However, major HCI conferences, such as CHI, and CSCW should consider making some of the open science practices a requirement for submission. This could include mandatory interview protocol sharing for qualitative works, codebook sharing, pre-registration of hypothesis-driven studies, and code and data sharing for quantitative research which are publicly funded projects. If the researchers are not able to share then they should provide proper justification for not sharing. Conferences could provide clear guidelines on what is expected and offer support to researchers unfamiliar with these processes with some webinars. Journals in the HCI field should also consider integrating open science guidelines\cite{silverstein2024guide} for the submission of supplementary materials, such as data and code, as part of the publication process. At the same time, we think pre-registration for qualitative/exploratory works should not be mandatory by these journals. 

\subsubsection{Incentivize Open Science Practices}
One of the most significant barriers to adopting open science practices identified in this study is the lack of clear and tangible incentives. Academic institutions should consider integrating open science contributions into the criteria for promotions, selection, and tenure\cite{schonbrodt2018academic}. By recognizing and rewarding efforts to share data, code, and research protocols openly, institutions can create a powerful incentive for researchers to adopt these practices\cite{knauer2024yes, merrett2021open}. Leading HCI conferences and journals should introduce specific recognition for papers that excel in open science practices. For example, awards could be given for best practices in data sharing, reproducibility, or the most comprehensive supplementary materials. Such recognition would not only motivate researchers but also set a benchmark for quality in science. Encourage the development and use of metrics that recognize the impact of open science practices\cite{heck20218, wilsdon2017next}, such as citations of datasets, code, and other shared materials. These metrics could be integrated into existing impact assessments, helping researchers see the tangible benefits of their open contributions.

\section{Limitations}
While our study provides important and valuable insights into the perceptions and challenges of adopting open science practices within the HCI community, several limitations should be acknowledged. The study is based on semi-structured interviews with 18 participants, which, while providing in-depth qualitative data, may not be fully representative of the broader HCI community. The perspectives captured in this study may vary from those of other researchers, particularly in different geographic regions or subfields of HCI. Our study primarily focuses on qualitative insights, exploring the attitudes, experiences, and perceived challenges related to open science practices. While this approach provides rich, contextual understanding, it does not offer quantitative measures of how widespread these attitudes and practices are within the HCI community. 

\section{Conclusion}
The adoption of open science practices within the HCI community presents both significant opportunities and challenges. Our research reveals that while there is a growing awareness of open science, its implementation remains unequal with different barriers. One of the main reasons for this is that researchers exhibit a wide range of understanding about open science practices. Key barriers to adoption include the lack of clear incentives and rewards, cultural resistance within the HCI community, limited training opportunities, concerns about intellectual property, limited time, and ethical concerns. These challenges are compounded by an academic culture that prioritizes traditional research outputs, such as publications and citations, over the principles of transparency and reproducibility central to open science. To advance the adoption of open science, our study recommends the development of institutional and conference-level incentives, the standardization of open science requirements, and the integration of comprehensive training programs into the education of HCI researchers.

\bibliographystyle{ACM-Reference-Format}
\bibliography{bibliography}


\end{document}